# The Universal Arrow of Time III-IV:(Part III) Nonquantum gravitation theory (Part IV) Quantum gravitation theory

## The Universal Arrow of Time III: Nonquantum gravitation theory.

### Kupervasser Oleg

### Abstract

The paper is dealing with the analysis of general relativity theory (theory of gravitation) from the point of view of thermodynamic time arrow. Within this framework «informational paradox» for black holes and «paradox with the grandfather» for time travel "wormholes" are resolved.

## 1. Introduction.

In paper we consider thermodynamic time arrow [1-2] (defined by a direction of the entropy increase) within the limits of not quantum relativistic gravitation theory. In the classical Hamilton mechanics any initial and final states are possible. Besides, between them there is a one-to-one correspondence. In relativistic theory of gravitation a situation is other. There are topological singularities of space which make possible a situation when for *finite* time different initial states give an identical final state. It is a collapse of black holes. On the other hand, having considered inverse process in time - white holes, we receive a situation when a single initial state can give a set of different final states for *finite* time. There are also situations of other sort - when not arbitrary initial states are possible. It is a case of "wormholes" through which it is possible to travel in the past. Thus there is necessary of self-consistency between the past and the future, making impossible some initial states. Black Holes lead to informational paradox, and "wormholes" - to «to paradox with the grandfather». Analysis of these situations with a point of view of thermodynamical time arrow and resolution of the defined above paradoxes are a topic of this paper.

## 2. Black Hole

In the modern cosmological models there are additional appearances, except the appearances already featured in the classical mechanics. In Einstein's relativity theory as well as in classical mechanics the motion is reversible. But there is also an important difference from the classical mechanics. It is *ambiguity* of a solution of an initial value problem: deriving a final state of a system from the complete set of initial and boundary conditions can give not single solution or no solution. In general relativity theory, unlike the classical mechanics, two various states for *finite* time can give infinitesimally close states. It happens at formation of a black hole as a result of a collapse. Hence, formation of the black hole goes with its entropy increase.

Let's consider an inverse process featuring a white hole. In this process infinitesimally close initial states for *finite* time can give different terminating states. Time reversion leads to appearing white hole and results in entropy decrease. The white hole can not exist in a reality because of the same reasons that processes with entropy decrease are impossible in the classical mechanics.

However its instability is much stronger than instability in the classical mechanics. It has finite value already with respect to *infinitesimally small* perturbations. As consequence there are alignment of thermodynamic time arrows between the white hole and the observer/environment. The white hole transforms to a black hole for the observer. It means that the observer/environment even *infinitesimally weakly* interacting with the white hole can affect considerably its evolution for finite time. Thus the gravitational interaction of the observer/environment with the white hole is always distinct from zero.

Here there is a well-known informational paradox [3]: the collapse leads to losses of the information in the Black Hole. It, in turn, results in incompleteness of our knowledge of a state of system and, hence, to unpredictability of dynamics of system, including Black Hole. The information, which in the classical mechanics always conserves, in a black hole disappears for ever. Is it really so? Or, probably, it is stored in some form inside of a black hole? Usually only two answers to this problem are considered: Or the information really vanishes completely; or the information is stored inside and can be extracted by some way. But, most likely, the third answer is true. Because of inevitable influence of the observer/environment it is impossible to distinguish these two situations experimentally in principle! And if it is impossible to verify something experimentally, it can not be a topic for the science.

Actually, suppose that the information is stored in a black hole. Whether is it possible to resolve informational paradox and to extract this information from it? Perhaps, we can reverse a collapsed black hole, to convert it into a white hole and to extract the disappeared information? It would seem impossibly. But recently there appeared an interesting paper, which though and not directly, but allows to make it [4]. There is proved, that a black hole is completely equivalent to an entry to the channel, pairing two Universes. The entry of this channel is similar to the black hole, and an exit is similar to the white hole. This white hole can be considered, in some sense, as the reversed black hole. But to verify that the information does not disappear, we should come into the second Universe. To do it we suppose, that there is some "wormhole" which connects these two Universes. Let the observer can pass it and observe the white hole. But even if it happens, we know that the white hole is extremely unstable with respect to any observation. Attempt to observe it will result in its transformation into a black hole. It will close any possibility to verify, that the information is stored. Hence, both solutions of informational paradox are really equivalent and observationally are not distinguishable.

This property of nonreversible information losses results in fact that the entropy increase law turns to be an exact law of the nature within framework of the gravitational theory. Really, there is such new fundamental value, as entropy of a black hole. It distinguishes gravitational theory from classical mechanics where the law entropy increase law has only approximate character (FAPP, for all practical purposes).

The accelerated expansion of the Universe results in the same effect of nonreversible information losses: there are unobservable fields, whence we are not reached even by light. Hence, these fields are unobservable, and the information stored in them is lost. It again results in unpredictability of relativistic dynamics.

## 3. Time wormhole

Let's consider from the point of view of the entropy such paradoxical object of general relativity theory, as time "wormhole" **[5]**. We will consider at the beginning the most popular variant, offered by Morris and Thorne **[6].** Let we have space wormhole with the extremities laying nearby. By very simple procedure (we will ship one of the extremities on a spaceship, we will move it with a velocity close to light, and then we will return this extremity on the former place) space wormhole can be conversed into time wormhole (wormhole traversing space into one traversing time). It can be used as a time machine. Such wormhole demands the special exotic matter necessary for conserving its equilibrium. However there were models of a time machine which allow to be bypassed absolutely without exotic substance [7, 9]. Or, using an electromagnetic field, allow to be bypassed by its small amount [8]. Use of such time machine can lead to well-known «paradox of the grandfather» when the grandson, being returned in the past, kills his grandfather. How this paradox can be resolved?

From the physical point of view, the paradox of the grandfather means, what not all initial states which exist before time machine formation are realizable. Introducing the additional feedback between the future and the past a time wormhole makes their impossible. Hence, we or should explain nonrealizability such initial states. Or suppose, that time "wormhole" is unstable, like a white hole, and easily changes.

Curiously enough, but both explanations are true. However for macroscopic whomeholes the first explanation has priority. Really, it would be desirable very much to have a macroscopic topology of the space to be stable. Constrains on initial states appears from entropy increase law and the correspondent alignment of thermodynamic time arrows, related to instability of states with opposite directions of these time arrows [1-2]. But macroscopic laws of thermodynamics are probability. For very small number of cases they are not correct (large-scale fluctuations). Both for these situations, and for microscopic wormholes where the concept of a thermodynamic time arrows and thermodynamics laws are not applicable, priority has the second explanation. It is related to extremal instability of the topology, which is defined by the time machine [9]. We discussed above such type of extremal instability for white holes. For macroscopic wormholes the solution can be discovered by means of the entropy increase law .It is ensured by instability of processes with the entropy decrease with respect to the Universe. This instability results in alignment of thermodynamic time arrows.

Really, space wormhole does not lead to paradox. The objects immersed by its one extremity, go out other extremity during later time. Thus, the objects from more ordered and low entropy past hit in less ordered and high entropy future. During a motion through the wormhole the entropy of the travelling objects also increases: they transfer from more ordered state in less ordered one. Thus, time arrow of the object, travelling inside of the wormhole, and time arrow of the world around the wormhole have the same directions. It is also true for travelling through the time wormhole from the past to the future.

However for travelling from the future to the past of the time arrow directions of the traveler into the wormhole and world around the wormhole will be already opposite [10, 11-13]. Really, the object travels from less ordered future to the more ordered past, but his entropy increases, instead of decreases! Hence, thermodynamic time arrows of the Universe and the traveler have opposite directions. Such process, at which entropies of the traveler decreases concerning the Universe, are unstable [1-2]. Hence, «memory about the past» of the traveler will be destroyed (and, may be, he will be destroyed completely), that will not allow him «to kill the grandfather».

Which mechanism at travelling in the wormhole ensures alignment of thermodynamic time arrows of the traveler and the Universe? Both extremities of "wormhole" it is the large bodies having finite temperature. Both extremities under the second thermodynamics laws inevitably should radiate light which partially hits to the wormhole. Already at the moment of "time machine" formation (transformation of the space wormhole into the time one) between its extremities there is a closed light ray. Every time when the ray features a circle, it is more and more biased to a violet part of the spectrum. Transiting a circle behind circle, rays are lost by the focal point, therefore energy does not amplify and it does not become infinite. Violet bias means, that the history of a particle of light is finite and defined by its coordinate time, despite the infinite number of circles [14]. This and other rays of light in wormhole fluctuate. They also have a direction of its thermodynamic time arrow coinciding with a thermodynamic time arrow of the Universe. Thanks to inevitable interaction with this radiation very unstable state of the traveler destroys. The state of the traveler is unstable because his thermodynamic time arrow is opposite to the Universe thermodynamic time arrows. The resulting destruction is enough to prevent the paradox of the grandfather.

"Free will" allows us to initiate freely only irreversible processes with the entropy increase, but not with its decrease. Thus, we cannot send a object from the future to the past. Process of alignment of thermodynamic time arrows and the correspondent entropy increase law forbids *the initial conditions* necessary for travelling of the macroscopic object to the past and resulting in "paradox of the grandfather".

In paper [10] it is strictly mathematically proved, that the thermodynamic time arrow cannot have identical orientation with the coordinate time arrow during all travel over closed timelike curve. Process of alignment of thermodynamic time arrows (related to instability of processes with entropy decrease) is such *physical mechanism* which actually ensures performance of the entropy increase law.

Macroscopic laws of thermodynamics are probabilistic. For very small number of cases they do not work (large-scale fluctuations). Both for these situations and for microscopic systems where thermodynamics laws are not applicable, the other explanation of the grandfather paradox have priority. In this case the time wormhole, like a white hole, appears unstable even with respect to infinitesimally weak perturbations from gravitation of travelling object. It can result in its fracture and prevention of the paradoxes, as is proved strictly in [9]. What are outcomes of reorganization of the space-time topology after fracture of the time wormhole? The author of [9] writes:

«As we argue … non-uniqueness does not let the time travel paradoxes into general relativity — whatever happens in a causal region, a space-time always can evolve so that to avoid any paradoxes (at the sacrifice of the time machine at a pinch). The resulting space-times sometimes … curiously remind one of the many-world pictures».

Let's formulate a final conclusion: *for macroscopic processes* instability of processes with the entropy decrease and correspondent alignment of thermodynamic time arrows makes almost impossible existence of initial conditions that allow travel to the past. Thereby it prevents both wormholes fracture and traveling of macroscopic bodies in the past leading to "paradox of the grandfather".

For very improbable situations of macroscopic wormholes and for microscopic wormholes the wormhole fracture must occur. This fracture is result of remarkable property of general relativity theory - extremal instability: infinitesimal external action (for example, gravitation from traveler) can produce wormhole fracture for finite time!

## 4. Conclusions.

Let's summarize above. Observation process should be taken into account inevitably during considering any physical process. We must transform from ideal dynamics over coordinate time arrow to observable dynamics with respect to thermodynamical time arrow of observer. It allows us to exclude all unobservable in the reality phenomena, leading to paradoxes. Thus it is necessary to consider following things. The observer inevitably is a nonequilibrium macroscopic chaotic body with the thermodynamic time arrow defined by his entropy increase direction. He yields all measurements with respect to this thermodynamic time arrow. Dynamics of bodies with respect to this thermodynamic time arrow is named as observable dynamics. It differs from ideal dynamics, with respect to the coordinate time arrow. All bodies are featured in observable dynamics in macroparameters, unlike the ideal dynamics using microparameters. The coordinate does not exist at thermodynamic equilibrium. It can change the direction and is not coincide with the coordinate time arrow of the ideal dynamics. Always there is a small interaction between the observer and observable system. It leads to alignment of thermodynamic time arrows of the observer and the observable systems.

We can see misterious situation. The same reasons which have allowed us to resolve paradoxes of wave packet reduction in quantum mechanics, paradoxes Loshmidt and Poincare in the classical mechanics allow to resolve informational paradox of black holes and the paradox of the grandfather for time wormholes. Remarkable universality!

# Acknowledgment

We thank Hrvoje Nikolic and Vinko Zlatic for discussions and debates which help very much during writing this paper.

# The Universal Arrow of Time IV: Quantum gravitation theory.

## Kupervasser Oleg

## Abstract

The paper is dealing with the analysis of quantum gravitation theory from the point of view of thermodynamic time arrow. Within this framework «informational paradox» for black holes and «paradox with the grandfather» for time travel "wormholes", black stars, Penrose's project of new quantum gravitation theory, anthropic principle are considered.

## 1. Introduction

The paper includes the analysis of quantum gravitation theory from the point of view of the thermodynamic time arrow [1-3]. Inside of this framework«informational paradox» for black holes and «paradox of the grandfather» for time "wormholes", black stars [4] and anthropic principle [5] are considered. It is shown, that wishes of Penrose [6-7] for the future theory of quantum gravitation need not creation of the new theory, and can be realized within a framework of already existing theories by the help of the thermodynamic approach.

## 2. Black holes

In general relativity theory, unlike the classical mechanics, two different states for *finite* time can give infinitesimally close states. It happens during formation of a black hole as a result of its collapse. It results in the well-known informational paradox [8]: the collapse leads to losses of the information in the black hole. It results in incompleteness of our knowledge of the system state. Hence, it can leads to unpredictability of the system dynamics. The information, which in classical and a quantum mechanics is always conserved, disappears in a black hole. Whether is it so? Usually only two answers to this problem are considered. Or the information really vanishes completely; or the information is conserved inside of the black hole and can be extracted. We will see that in quantum gravitation we have the same answer, as in general relativity theory - both answers are possible and true, because the difference is not observed experimentally.

For the semiclassical theory of gravitation where gravitation is featured by relativistic relativity theory, and fields is featured by quantum field theory, the paradox resolution is made with the help of Hawking radiation.

In quantum field theory the physical vacuum is filled by permanently born and disappearing "virtual particles". Close to the event horizon (but nevertheless outside) of a black hole the pairs of particle-antiparticle can be born from vacuum. It is possible the situation when an antiparticle total energy appears subzero, and a particle total energy appears plus. Falling to the black hole, the antiparticle reduces its total energy and mass while the particle is capable to fly away to infinity. For the remote observer it looks as Hawking radiation of the black hole.

Since this radiation incoherent after evaporation of the black hole all information accumulated inside of it disappears. It is an answer of the semiclassical theory. It would seem that this result contradicts to reversibility and unitarily of quantum mechanics where the information can not be lost. We would expect the same result in quantum gravitation theory. But whether is it so?

We have now no finished theory of quantum gravitation. However for a special case of the 5-dimentional anti-de-Sitter space this paradox is considered by many scientists to be resolved. The information is supposed to be conserved, because a hypothesis about AdS/CFT dualities, i.e. hypotheses that quantum gravitation in the 5-dimensional anti-de-Sitter space (that is with the negative cosmological term) is equivalent mathematically to a conformal field theory on a 4-surface of this world [9]. It was checked in some special cases, but not proved yet in a general case.

Suppose that if this hypothesis is really true, as it automatically solves the information problem. The fact of the matter is that the conformal field theory is unitary. If it is really dual to quantum gravitation then the corresponding quantum gravitation theory is unitary too. So, the information in this case is not lost.

Let's note that it not so. Taking into account the influence of the observer makes inevitable information losses. Process of black hole formation and its subsequent evaporation happens on all surface of the anti-de-Sitter world (described by the conformal quantum theory) which includes as well the observer. The observer inevitably gravitational interacts with the black hole and its radiation. Unlike usual quantum mechanics all-pervading gravitational interaction exists in quantum gravitation. So influence of the observer already cannot be made arbitrary small under any requirements. Interaction with the observer makes the system not unitary similarly to the semiclassical case.

It would seem that we can solve the problem by including the observer in the system description. But the observer cannot precisely know the initial state and analyze the system when he is its part! So, he cannot experimentally verify the difference between unitary and not unitary evolution. It is necessary complete knowledge of the system state for such verification. But it is impossible at introspection.

In the anti-de-Sitter world Universe expansion is inevitably replaced by a collapse. But the same effect information losses are available also for the accelerated expansion of the Universe - there appear unobservable parts of Universe, whence we are not reached even by light. Hence, these parts are unobservable, and the information containing in them is lost. It again results in unpredictability.

Thus, the experimental verification of informational paradox again becomes impossible *in principle*! In case of quantum gravitation information conservation happens only on a paper in the ideal dynamics. In the real observable dynamics the difference is not observed experimentally in principle. It is possible to consider both answers to the problem to be correct. The two cases of conservation or not conservation of information experimentally are not distinguishable.

Principal difference between usual quantum theory and quantum gravitation theory happens because of inevitable gravitational interaction. In usual quantum theory interaction between an observer and an observed system can be made zero in principle at known initial conditions of the observed system. In quantum gravitational systems the small gravitational interaction with the observer is irremovable in principle: it creates principally inherent decoherence and converts evolution of any observable system into the nonunitary. Only for nonobservable ideal evolution on a paper it can be made formally unitary. But also it is possible not to make it unitary- here we have a freedom to choose. If we wish to feature real observable dynamics we can put the dynamics to be nonunitary. For macrobodies such observable dynamics is semiclassical theory. It is experimentally indistinguishable for the real macroscopic observer from unitary quantum gravitation dynamics of large black holes.

## 3. Time wormhole

Let's consider from the point of view of the entropy such paradoxical object of general relativity theory, as time "wormhole" **[10]**. We will consider at the beginning the most popular variant, offered by Morris and Thorne **[11].** Let we have space wormhole with the extremities laying nearby. By very simple procedure (we will ship one of the extremities on a spaceship, we will move it with a velocity close to light, and then we will return this extremity on the former place) space wormhole can be conversed into time wormhole (wormhole traversing space into one traversing time). It can be used as a time machine. Such wormhole demands the special exotic matter necessary for conserving its equilibrium. However there were models of a time machine which allow to be bypassed absolutely without exotic substance [12, 17]. Or, using an electromagnetic field, allow to be bypassed by its small amount [13]. Use of such time machine can lead to well-known «paradox of the grandfather» when the grandson, being returned in the past, kills his grandfather. How this paradox can be resolved?

Let's consider, what answer to this problem the semiclassical theory of gravitation gives. Suppose that the macroscopic topology of the space related to the time machine is unchanged. At the moment of the time machine formation (transformation of the space wormhole into time one) between its extremities there is a closed light ray. Its energy does not reach infinity, despite the infinite number of passes, because of a defocusing of the light [16]. Other situation however arises in the semiclassical theory with «vacuum fluctuations» radiation field [14]. Passing the infinite number of times through the wormhole and summing, these fluctuations reach the infinite energy which will destroy any traveler.

However a situation in quantum gravitation is the other. Quantum fluctuations contain large energies when they arise on short distances. So it is possible to find so small distance on which energy of fluctuation will be large enough for formation of a tiny black hole. Thus the horizon of the tiny black hole will be the same size, as well as this small distance. The space - time is not capable to remain homogeneous on such short distances. This mechanism ensures natural "blocking" of singular fluctuations formation, restricting them on a size - «maximum energy in the minimal sizes» [16].

Detailed calculations of quantum gravitation show [15], that this "blocking" to formation of singular fluctuations ensures very small, but not zero probability of unobstructed transiting of time "wormhole" for macroscopic object. How to prevent in this situation «paradox of the grandfather»? Here it is convenient for us to use the language of multiworld interpretation of quantum mechanics. To prevent paradox the traveller should penetrate to the parallel world where it can easily «kill the grandfather» without breaking a causality principle. Such parallel world will interfere quantum-mechanically with the worlds of "not killed grandfather" where the observer was unsuccessful to transit the time wormhole. However the probability amplitude of such world will be extremely small. Whether can the observer in the world where «the grandfather has not killed» discover the alternative world at least in principle, using quantum correlations between the worlds? Similarly to "paradox of the Schrodinger cat" he cannot because of the same reasons, as in usual quantum mechanics [2]. Observation of large effects of quantum correlations is impossible because of «observer's memory erasing» [1-2]. Penetration to the parallel world of quantum mechanics is experimentally indistinguishable from time wormhole fracture and penetration to the parallel world of general relativity theory [3, 17]. It means, that from the point of view of the external real macroscopic observer a situation when the traveler has perished in the wormhole or has penetrated in «other world», are observationally indistinguishable. It is equivalent to a situation when the traveler falls into a black hole. we do not know - whether he is crushed in the singularity or penetrated into «other world» through the white hole [18]. (Though this difference is observed and essential for the traveler. But he will carry away all these observations with himself into «other world».) We see, that as well as in a case of "informational paradox", the difference between quantum and semiclassical theories for macroscopic objects experimentally is not observed for the macroscopic observer which did not travel in the time wormhole.

## 3. Black stars.

Recently there appears an interesting theory of "black stars" [4]. Usually the collapse of a black hole is considered as fast process. However we don't know well a matter behavior under high pressures. We know that intermediate stages such as white dwarfs, neutron stars are possible before a black hole collapse. These intermediate stages make a collapse not avalanche, but gradual. Probably, on the way to a collapse will appear additional intermediate stages appear, for example, quark stars. These intermediate stages make this process to be gradual without a fast collapse at all. For classical gravitation it is incidental. The star becomes a black hole for gradual process also. But for semiclassical gravitation it is important. It can be shown that for such case at slow squeezing quantum fluctuations at a surface will prevent a star material to collapse to a singularity and to become a black hole. Outside this object would be similar to a black hole, but inside it would be different, conserving all information without singularity. It will allow for a traveler to penetrate its surface and to come back. It is worthy of note that against such picture there is an essential objection.

How much is such construction of a star stable with respect to the external perturbation imported by the traveler to inside of a star? Also how much is the traveler stable during such travel? The traveler is a macroscopic body. After penetration to a black star, he will increase its mass stepwise at finite value. It can results in its collapse to a black hole. Suppose that the process again goes "gradually" without collapse. Then the traveler "would be dissolved" into the star and cannot come back also. Thus, it seems that the difference between a black star and a black hole can not observe experimentally. So, it means that the difference between these objects exists only on a paper, i.e. in ideal dynamics.

## 4. Penrose's project of new quantum gravity theory.

In the nice books [6-7] Penrose gives the remarkable prediction of the future theory of quantum gravitation. In this theory:

1) Unlike usual quantum mechanics a wave packet reduction is fundamental property of the theory.
2) This reduction inseparably linked with gravitation appearance.
3) The reduction is ruled not only by probability law. It can be result in some more complex uncertain behavior that can not be predicted even by probability law.
4) Unlike remarkable coherent quantum systems, classical chaotic nonequilibrium systems are exposed to criticism. They are supposed to be not relevant for modeling of real complex systems. Describes above unpredictable systems must be only pure quantum system.

It is worthy of note, that for receiving all these properties we need not new theory. Let take into account an inevitable gravitational interaction of the macroscopic real observer and his thermodynamical time arrow. It results in all described above outcomes within framework of already existing theories of quantum gravitation. Besides, classical chaotic nonequilibrium systems possess all properties of quantum ones. For any «purely quantum effect» always it is possible to discover such classical analogue (Appendix A [2]). Namely:

1) We saw above, that an inevitable gravitational interaction of the macroscopic real observer with an unstable observable system inevitably makes evolution of the observable system nonunitary. The difference between the unitary and nonunitary theory exists only on a paper and experimentally is not observed in quantum gravitation theory.
2) Because of the reasons stated above the gravitation interaction results in the inevitable reduction and correspondent nonunitarity in framework of the current quantum gravitation theory. Moreover, for macroscopic objects the semiclassical theory is already possessing desirable fundamental property of nonunitary. It is experementally equvalent to the quantum gravitation theory.
3) Behavior of many macroscopic bodies, in spite of nonunitarity, it is possible to describe completely by set of macroparameters and laws of their evolution. There are, however, *unpredictable* systems, whose behavior can not be described completely even by probability laws.
   For example, let us consider quantum computers. Suppose that some person started such quantum computer and knows its initial state. Its behavior is completely predicted by such person. However for the second person who is not present at start, its behavior is *uncertain* and *unpredictable*. Moreover, an attempt of the second person to observe some intermediate state of the quantum computer would result in destroying its normal operation.
   In case of quantum gravitation even the person started quantum computer cannot predict its behavior. Indeed the inevitable gravitational interaction between the person and the quantum computer will make such prediction impossible. Thus, «the unpredictability which is distinct from a probability law» becomes fundamental property of any quantum gravitation theory.
4) Unstable classical systems in many aspects remind on the properties of the quantum system (Appendix A [2]). Moreover, mathematical models of classical analogues of quantum computers exists [19]. Some paradoxical properties of the life objects reminding quantum computers can be modeled by classical unstable systems [20].

Summing up, we can see that all wishes of Penrose are realizable within the framework of the existing paradigm and need not any new fundamental theory. Moreover, all properties of macroobjects are usually described by macroparameters to exclude influence of the macroscopic observer. That inevitably results in unobservability too small intervals of time and space. So it is possible to construct their observable dynamics on basis of "discrete model of space-time". But such dynamics would not be a new theory. For any macroscopic observer the dynamics would be experimentally indistinguishable from the current quantum theory of gravitation.

# 5. Anthropic principle in quantum gravity theory.

The number of the possible vacuum states in quantum gravitation theory is equal to a very large value. For a selection of suitable vacuums usually anthropic principle is used [5]. It means that system evolution should results in appearance of an observer which is capable to observe the Universe. But such formulation has too philosophical character. It is complex to use it in practice. We can formulate here more accurate physical principles which are equivalent to the anthropic principle:

The initial state of Universe should result in formation of its substance in the form of a set of many macroscopic nonequilibrium objects weakly interacting with each other. These objects should have entropy and temperature. They should have thermodynamic time arrows. Small local interaction between objects should results in alignment of thermodynamic time arrows. Though these objects consist of many particles and are described by a huge set of microparameters, evolution of these objects can be described by a set of macroparameters, except for rare instable state.

However these unstable states play a important role, forming a basis for origin of a observer in the Universe. There should be unstable global correlations between Universe parts, and nonequilibrium macrosystems with local interior correlations which are origin of the observer.

We can conclude. To get the situation described above, the initial state of the Universe should be highly ordered and has the low entropy.

I.e., in short, evolution should results in the world that can be described in the thermodynamical form [1-3, 21-23]. Only such world can be origin of an observer, which is capable to study this world.

# 6. Conclusions.

We see that the informational paradox and the paradox of the grandfather are resolved in the quantum gravitational theory very similarly to the nonquantum general relativity theory. It is realized by consideration of weak interaction of systems with the real nonequilibrium macroscopic observer. Moreover, this approach (similarly to usual quantum theory) allows to resolve the wave packet reduction problem. But this reduction in quantum gravitation becomes fundamental property of the theory, unlike usual quantum mechanics. Such approach allows to consider the other complicated questions of quantum gravitation - anthropic principle, black stars.

# Acknowledgment

We thank Hrvoje Nikolic и Vinko Zlatic for discussions и debates which help very much during writing this paper.

# Универсальная стрела времени III:
# Неквантовая гравитационная теория.

Купервассер О.Ю.

**Аннотация.**

Статья посвящена анализу общей теории относительности (гравитации) с точки зрения термодинамической стрелы времени. В рамках этого рассмотрения разрешены «информационный парадокс» для черных дыр и «парадокс с дедушкой» для временных «червоточин».

## Введение

В статье мы рассмотрим термодинамическую стрелу времени [1-2] (определяемую направлением роста энтропии) в рамках неквантовой релятивисткой теории гравитации. В классической гамильтоновой механике любые начальные и конечные состояния возможны. Кроме того, между ними существует взаимно-однозначное соответствие. В релятивистской теории гравитации ситуация иная. Имеются топологические особенности пространства, которые делают возможным ситуацию, когда за *конечное* время разные начальные состояния дают одинаковое конечное состояние. Это коллапс черных дыр. С другой стороны, рассмотрев обратный во времени процесс – белые дыры, мы получим ситуацию, когда одному начальному состоянию за *конечное* время соответствуют разные конечные состояния. Имеются и ситуации другого сорта – когда не любые начальные состояния возможны. Это случай «червоточин», через которые возможно путешествия в прошлое. При этом становится необходимым дополнительное само-согласование прошлого и будущего, делающее невозможным некоторые начальные состояния. Черные Дыры приводят к информационному парадоксу, а «червоточины» - к «парадоксу с дедушкой». Рассмотрению этих особых ситуаций релятивисткой теории гравитации с точки зрения термодинамической стрелы времени и разрешению связанных с ними парадоксов посвящена данная статья.

## 1. Черные дыры

В современных космологических моделях есть дополнительные явления, кроме явлений, уже описанных в классической механике. В общей теории относительности Эйнштейна движение так же, как и в классической механике обратимо. Но имеется и важное отличие от классической механики. Это *неоднозначность* решения задачи Коши: получения конечного состояния системы из полного набора начальных и граничных условий. В общей теории относительности, в отличие от классической механики, два различных состояния за *конечное* время могут дать бесконечно близкие состояния. Это происходит при образовании черной дыры в результате коллапса. Следовательно, образование черной дыры идет с увеличением энтропии.

Рассмотрим обратный процесс, описывающий белую дыру. В этом процессе бесконечно близкие начальные состояния за *конечное* время могут дать разные конечные состояния. Обращение времени приводит к появлению белой дыры и ведет к уменьшению энтропии. Белая дыра не может существовать в реальности по тем же причинам, что невозможны процессы с уменьшением энтропии в классической механике. Однако, ее неустойчивость намного более сильная, чем в классической механике. Она возникает уже по отношению к *бесконечно* малым возмущениям. Как следствие  возникает синхронизации собственных стрел времени белой дыры и наблюдателя/окружения. Белая дыра превращается для наблюдателя в черную дыру.
Это означает, что наблюдатель/окружение, даже бесконечно слабо взаимодействующий с белой дырой может значительно повлиять на ее эволюцию за конечное время. При этом гравитационное взаимодействие наблюдателя/окружения с белой дырой всегда отлично от нуля.

Здесь возникает знаменитый информационный парадокс [3]: Коллапс приводит к потере информации в Черной Дыре. Это, в свою очередь, ведёт к неполноте нашего знания о состоянии системы и, следовательно, к непредсказуемости динамики системы, ее включающей. Информация, которая в классической механике всегда сохраняется, в черной дыре исчезает навсегда. Так ли это? Или, возможно, внутри черной дыры она храниться в какой-либо форме? Обычно рассматривают только два ответа на этот вопрос. Либо информация действительно пропадает бесследно; либо информация сохраняется внутри нее и может быть каким-то путем извлечена. Но, скорее всего, верным является третий ответ. Из-за неизбежного влияния  наблюдателя/окружения экспериментально различить эти две ситуации просто невозможно! А что нельзя проверить экспериментально, не должно является предметом науки и обсуждения.

На самом деле, предположим, что информация сохраняется внутри черной дыры. Можно ли разрешить информационный парадокс и извлечь эту информацию из нее? Может быть, мы можем обратить каким либо образом сколлапсировавшую черную дыру, превратить ее в белую дыру, и извлечь исчезнувшую информацию? Казалось бы, это невозможно. Но недавно появилась интересная работа, которая, хоть и не напрямую,  но позволяет сделать это  [4]. В ней доказывается, что черная дыра полностью эквивалентна входу в канал, соединяющий две Вселенные. Причем вход этого канала подобен черной дыре, а выход белой. Эта белая дыра и может рассматриваться, в некотором смысле как обращенная черная дыра. Но для того, чтобы убедиться, что информация не исчезает, мы должны проникнуть во вторую Вселенную. Предположим, что существует некая «червоточина», которая соединяет две Вселенные. Пусть наблюдатель может проникнуть через нее и пронаблюдать за белой дырой. Но даже если это случится, мы знаем, что белая дыра неустойчива по отношению к наблюдению. Попытка ее наблюдения приведет к её превращению в черную дыру. Это закроет всякую возможность подтвердить, что информация сохраняется.  Следовательно, оба решения информационного парадокса действительно равноправны и экспериментально не различимы.

Это свойство необратимой потери  информации приводит к тому, что закон возрастания энтропии превращается в точный закон природы в рамках гравитационной теории.  Действительно, появляется такая новая фундаментальная величина, как энтропия черной дыры. Это отличает гравитационную теорию от классической механики, где  закон возрастания энтропии носит лишь приближенный характер (FAPP, для всех практических целей).

Тот же эффект необратимой потери информации имеет и ускоренное расширение Вселенной – появляются ненаблюдаемые области, откуда до нас не доходит даже свет. Следовательно, они ненаблюдаемые, и содержащаяся в них информация потеряна. Это опять ведет к непредсказуемости релятивисткой динамики.

## 2. Временная червоточина.

Рассмотрим с точки зрения энтропии и такой парадоксальный объект общей теории относительности, как временная «червоточина» (кротовая нора) [5]. Рассмотрим вначале ее наиболее популярный вариант, предложенный Моррисом и Торном [6]. Пусть у нас имеется пространственная кротовая нора с лежащими рядом  концами. Путем очень простой процедуры (погрузим одного из концов на космический корабль, обеспечим его движение со скоростью сравнимой со световой, а затем вернем этот конец на прежнее место) пространственная кротовая нора может быть преобразована во временную (wormhole traversing space into one traversing time). Она может быть использована как машина времени. Подобная кротовая нора требует особого экзотического вещества, необходимого для поддержания ее равновесия. Однако есть модели машины времени, которые или позволяют обойтись совсем без экзотического вещества [7, 9]. Или же, используя электромагнитное поле, позволяют обойтись его малым количеством [8]. Использование этой машины времени может приводить к знаменитому «парадоксу дедушки», когда внук, возвращаясь в прошлое, убивает своего дедушку. Как же может быть разрешен этот парадокс?

С физической точки зрения, парадокс дедушки означает, что не все начальные состояния, которые существуют до образования машины времени осуществимы. Дополнительная обратная связь между будущим и прошлым через временную червоточину делает их невозможными. Следовательно, мы либо должны объяснить нереализуемость таких начальных состояний, либо допустить, что временная «червоточина» неустойчива, наподобие белой дыры, и легко разрушается.

Как ни странно, оба объяснения в принципе верны. Однако для макроскопических червоточин приоритетным является первое объяснение. Действительно, очень хотелось бы иметь макроскопическую топологию пространства стабильной. Ограничение на начальные состояния при этом связано с законом роста энтропии и синхронизацией термодинамических стрел времени, связанной с неустойчивостью состояний с разной направленностью этих временных стрел [1-2]. Макроскопические законы термодинамики вероятностны. Для очень небольшого числа случаев они не действуют (крупномасштабные флюктуации). Как для этих ситуаций, так и для микроскопических червоточин, где понятие термодинамической стрелы времени и законы термодинамики не применимы, приоритетным оказывается второе объяснение. Оно связанно с экстремальной неустойчивостью топологии, определяемой машиной времени [9], аналогичной неустойчивости белой дыры. Для макроскопических кротовых нор разрешение может быть найдено с помощью закона возрастания энтропии, обеспечиваемого неустойчивостью процессов с убыванием энтропии относительно Вселенной и вытекающей из этого синхронизацией термодинамических стрел времени.

Действительно, пространственная кротовая дыра не приводит к парадоксу. Объекты, поглощенные ее одним концом, выходят из другого конца в более позднее время. Таким образом, объекты из более упорядоченного низкоэнтропийного прошлого попадают в менее упорядоченное высокоэнтропийные будущее. При движении вдоль кротовой норы энтропия путешествующих объектов также растет: они переходят из более упорядоченного состояния в менее упорядоченное. Таким образом, собственные стрелы времени путешествующего в кротовой норе объекта и окружающего мира сонаправлены. Тоже верно для путешествия по временной кротовой норе из прошлого в будущее.

Однако для путешествия из будущего в прошлое стрелы времени путешественника в кротовой норе и окружающего мира будут уже противоположны [10, 11-13 ]. Действительно, сам объект путешествует из менее упорядоченного будущего в более упорядоченное прошлое, но при этом его собственная энтропия растет, а не убывает! Следовательно, термодинамические стрелы времени Вселенной и путешественника разнонаправлены. Такой процесс, при котором энтропии путешественника убывает относительно Вселенной, неустойчив [1-2]. Следовательно, «память о прошлом» путешественника (а, может, и он сам полностью) будет разрушена, что не позволит ему «убить дедушку».

Какой именно механизм при путешествии в кротовой норе обеспечивает синхронизацию стрел времени путешественника и Вселенной? Оба конца «червоточины» это массивные тела, имеющие конечную температуру. Оба эти конца по законам термодинамики неизбежно должны излучать свет, который частично попадает и в кротовую нору. Уже в момент образования «машины времени» (преобразования пространственной червоточины во временную) между ее концами появляется замкнутый световой луч. Всякий раз, когда луч описывает окружность, он все больше смещается к фиолетовой части спектра. Проходя круг за кругом, лучи теряют фокус, поэтому энергия не усиливается и не становится бесконечной. Фиолетовое смещение означает, что история частицы света конечна и определена ее собственным координатным временем, несмотря на бесконечное число кругов [14]. Этот и иные потоки света в кротовой норе флюктуируют и имеют направление термодинамической стрелы времени, совпадающего с термодинамической стрелой времени Вселенной. Благодаря неизбежному взаимодействию с этими излучением разрушается очень неустойчивое состояние путешественника, имеющего обратное по отношению к Вселенной направление собственного термодинамического времени. Это разрушение происходит до степени, достаточной для предотвращения парадокса дедушки.

«Свобода воли» позволяет нам свободно инициировать лишь устойчивые процессы с ростом энтропии, но не с ее убыванием. Таким образом, мы не сможем послать объект из будущего в прошлое. Процесс синхронизации стрел времени и вытекающий из него закон роста энтропии запрещает *начальные условия*, необходимые для путешествия макроскопических объектов в прошлое и реализацию «парадокса дедушки».

В работе [10] строго математически доказывается, что собственная термодинамическая стрела времени не может все время иметь одинаковую ориентацию с собственной координатной стрелой времени при путешествии по замкнутой временноподобной траектории (closed timelike curve). Процесс синхронизации стрел времени (связанный с неустойчивостью процессов с убыванием энтропии) является тем самым *физическим механизмом*, который фактически обеспечивает выполнение закона роста энтропии.

Макроскопические законы термодинамики вероятностны. Для очень небольшого числа случаев они не действуют (крупномасштабные флюктуации). Как для этих ситуаций, так и для микроскопических систем, где законы термодинамики не применимы, приоритетным оказывается другое объяснение парадокса дедушки. Существуют два процесса: весь Космос и объект, путешествующий по червоточине из будущего Космоса в его прошлое. При этом временная червоточина, подобно белой дыре, оказывается неустойчивой даже по отношению к бесконечно малым возмущениям от гравитации путешествующего объекта, что может привести к ее разрушению и предотвращению парадоксов, что и доказывается строго в [9]. Каковы результаты перестройки топологии пространства-времени после разрушения временной червоточины? Автор [9] пишет:
«Как мы объясняли … неоднозначность, не позволяет существование парадоксов путешествия во времени в общей теории относительности - независимо от того, что произошло бы в причинной области, пространство-время всегда может развиваться так, чтобы избежать любых парадоксов (жертвую машиной времени, в крайнем случае). Получающееся при этом пространство-время иногда … любопытно напоминает одну из много-мировых картин.»

Следует отметить, что с точки зрения внешнего реального макроскопического наблюдателя ситуация, когда путешественник погиб в червоточине или попал в «иной мир», экспериментально неотличимы. Это эквивалентно ситуации, когда путешественник падает в черную дыру: нам не известно будет ли он раздавлен в сингулярности или попадет в «иной мир» через белую дыру. (Хотя для самого путешественника эта разница наблюдаема и существенна. Но он унесет свои все эти свои наблюдения с собой в «иной мир».)

Сформулируем окончательный вывод: *для макроскопических процессов* неустойчивость процессов с убыванием энтропии и сопутствующая ей синхронизация стрел времени в подавляющем числе случаев делает невозможным появление начальных условий несовместимых с существованием заданных червоточин. Тем самым предотвращается как их разрушение, так и путешествия по ним макроскопических тел в прошлое, приводящее к «парадоксу дедушки».

Для очень маловероятных ситуаций в случае макрообъектов и для микроскопических систем может реализоваться уже ранее отмеченное замечательное свойство экстремальной неустойчивости общей теории относительности: бесконечно малое внешнее воздействие может повлечь разрушение червоточины за конечное время!

## Выводы.

Подведем общие итоги. Процесс наблюдения должен неизбежно учитываться при рассмотрении всех физических процессов, чтобы исключить появление ненаблюдаемых в реальности явлений, приводящих к парадоксам. При этом нужно учитывать следующие вещи. Наблюдатель неизбежно является неравновесным макроскопическим хаотическим телом с термодинамической стрелой времени, определяемой направлением роста энтропии. Все измерения он производит относительно этой временной стрелы времени. Динамика тел, относительно этой стрелы времени называется наблюдаемой динамикой и отличается от идеальной динамики, относительно координатной стрелы времени. Все тела описываются в наблюдаемой динамике макропараметрами, в отличие от идеальной динамики, использующей микропараметры. Термодинамическая стрела времени не существует при термодинамическом равновесии. Она может менять свое направление и не совпадать с координатной стрелой времени идеальной динамики. Всегда существует малое взаимодействие между наблюдателем и наблюдаемой системой. Оно приводит к синхронизации термодинамических стрел времени наблюдателя и наблюдаемой системы.

Мы видим необыкновенную вещь. Все эти соображения, которые нам позволили разрешить парадокс редукции в квантовой механике, парадоксы Лошмидта (Loshmidt) и Пуанкаре в классической механике позволяют разрешить информационный парадокс черных дыр и парадокс дедушки для временных кротовых нор. Замечательная универсальность!

## Благодарности

## Библиография

# Универсальная стрела времени IV: Квантовая теория гравитации.

Купервассер О.Ю.

**Аннотация.**
Проанализированы парадоксы и проблемы квантовой теории гравитации с точки зрения термодинамического подхода.

## 1. Введение

Статья посвящена анализу квантовой теории гравитации с точки зрения термодинамической стрелы времени [1-3]. В рамках этого рассмотрения разрешены «информационный парадокс» для черных дыр и «парадокс с дедушкой» для временных «червоточин», рассмотрены черные звезды [4] и антропный принцип [5]. Показано, что пожелания Пенроуза [6-7] к будущей теории квантовой гравитации не требуют создания новой теории, а реализуемы в рамках уже существующих теорий при учете термодинамического подхода.

## 2. Черные дыры

В общей теории относительности, в отличие от классической механики, два различных состояния за *конечное* время могут дать бесконечно близкие состояния. Это происходит при образовании черной дыры в результате коллапса. За счет этого возникает знаменитый информационный парадокс [8]: коллапс приводит к потере информации в Черной Дыре. Это, в свою очередь, ведёт к неполноте нашего знания о состоянии системы и, следовательно, к непредсказуемости динамики системы, ее включающей. Информация, которая в классической и квантовой механике всегда сохраняется, в черной дыре исчезает. Так ли это? Обычно рассматривают только два ответа на этот вопрос. Либо информация действительно пропадает бесследно; либо информация сохраняется внутри нее и может быть каким-то путем извлечена. Мы увидим, что в квантовой гравитации ответ тот же, что и для общей теории относительности – оба ответа возможны и верны, поскольку разница экспериментально не наблюдаема.

Для квазиклассической теории гравитации, где гравитация описывается общей теорией относительности, а поля - квантовой теорией, разрешение парадокса находится с помощью излучение Хокинга.

В квантовой теории поля физический вакуум наполнен постоянно рождающимися и исчезающими «виртуальными частицами». Вблизи (но всё же снаружи) горизонта событий чёрной дыры прямо из вакуума могут рождаться пары частица-античастица. При этом возможен случай, когда полная энергия античастицы оказывается отрицательной, а полная энергия частицы - положительной. Падая в чёрную дыру, античастица уменьшает её полную энергию покоя, а значит, и массу, в то время как частица оказывается способной улететь в бесконечность. Для удалённого наблюдателя это выглядит как излучение Хокинга чёрной дыры.

Поскольку это излучение некогерентное после испарения черной дыры вся запасенная в ней информация исчезает – это ответ квазиклассической теории. Казалось бы, это противоречит обратимости и унитарности квантовой механики, где информация не теряется. Того же мы ожидали бы от квантовой теории гравитации. Но так ли это?

Мы не имеем сейчас законченную теорию квантовой гравитации. Однако для частного случая 5 мерного анти-де-Ситтеровского мира этот парадокс ныне многими учеными считается разрешенным в пользу сохранения информации, вследствие гипотезы о AdS/CFT дуальности, т. е. гипотезы о том, что квантовая гравитация в анти-де-ситтеровском (то есть с отрицательным космологическим членом) 5-мерном пространстве математически эквивалентна конформной теории поля на 4-поверхности этого мира. [9] Она была проверена в некоторых частных случаях, но пока не доказана в общем виде.

Полагают, что если эта гипотеза действительно верна, то это автоматически влечёт за собой разрешение проблемы об исчезновении информации. Дело в том, что конформная теория поля по построению унитарна. Если она дуальна квантовой гравитации, то значит и соответствующая квантово-гравитационная теория тоже унитарна, а значит, информация в этом случае не теряется.

Отметим, что это не так. Учет влияния наблюдателя делает неизбежной потерю информации. Процесс образования черной дыры и ее дальнейшее испарения происходит на всей поверхности анти-де-Ситтеровского мира (описываемого квантовой теорией поля), который включает также и наблюдателя. Наблюдатель неизбежно гравитационно взаимодействует с черной дырой и ее излучением. В отличие от обычной квантовой механики и по причине этого всепроникающего гравитационного взаимодействия, влияние наблюдателя теперь уже нельзя сделать пренебрежимо малым ни при каких условиях. Взаимодействие с наблюдателем делает систему не унитарной.

Казалось бы, мы можем включить наблюдателя в описание системы. Но наблюдатель не может точно знать свое начальное состояние и анализировать систему, частью которой он сам и является! А значит, не может проверить экспериментально разницу между унитарной и не унитарной эволюцией. Для этого необходимо знание полного состояния системы, что невозможно при самонаблюдении.

В анти-де-Ситтеровского мире расширение Вселенной неизбежно сменяется сжатием. Но тот же эффект потери информации имеется и при ускоренном расширении Вселенной – появляются ненаблюдаемые области, откуда до нас не доходит даже свет. Следовательно, они ненаблюдаемые, и содержащаяся в них информация потеряна. Это опять ведет к непредсказуемости.

Таким образом, экспериментальная проверка информационного парадокса снова становится невозможной *в принципе*! В случае квантовой гравитации сохранение информации происходит лишь на бумаге в идеальной динамике. В реальной наблюдаемой динамике разница не наблюдаема экспериментально в принципе. Оба ответа на вопрос о сохранении или не сохранении информации можно считать приемлемыми, поскольку они экспериментально не различимы.

Главное отличие столь большой разницы между обычной квантовой теорией и квантовой теорией гравитации происходит из того, что взаимодействие наблюдателя в обычной квантовой теории можно сделать нулевым в принципе при известных начальных условиях. В квантовых гравитационных системах малое гравитационное взаимодействие с наблюдателем неустранимо в принципе – что делает в принципе неустранимой декогеренцию и превращает эволюцию любой наблюдаемой системы в неунитарную. Лишь для непроверяемой идеальной эволюции на бумаге ее можно сделать формально унитарной. А можно и не делать – здесь у нас свобода выбора. Если мы хотим описывать реальную наблюдаемую динамику – делать этого не стоит. Для массивных тел такая наблюдаемая динамика - это квазиклассическая теория, которая экспериментально неотличима для реального макроскопического наблюдателя от квантовой гравитации массивных гравитационных черных дыр.

## 3. Временная червоточина.

Рассмотрим с точки зрения энтропии и такой парадоксальный объект общей теории относительности, как временная «червоточина» (кротовая нора) [10]. Рассмотрим вначале ее наиболее популярный вариант, предложенный Торном [11]. Пусть у нас имеется пространственная кротовая нора с лежащими рядом концами. Путем очень простой процедуры (погрузим одного из концов на космический корабль, обеспечим его движение со скоростью сравнимой со световой, а затем вернем этот конец на прежнее место) пространственная кротовая нора может быть преобразована во временную (wormhole traversing space into one traversing time). Она может быть использована как машина времени. Подобная кротовая нора требует особого экзотического вещества, необходимого для поддержания ее равновесия. Однако недавно появились новые модели машины времени, которые или позволяют обойтись совсем без экзотического вещества [12,17]. Или же, используя электромагнитное поле, позволяют обойтись его малым количеством [13]. Использование этой машины времени может приводить к знаменитому «парадоксу дедушки», когда внук, возвращаясь в прошлое, убивает своего дедушку. Как же может быть разрешен этот парадокс?

Рассмотрим, какой ответ на этот вопрос дает квазиклассическая теория гравитации, где гравитация описывается общей теорией относительности, а поля – квантовой теорией. Мы при этом полагаем макроскопическую топологию пространства, связанную с машиной времени неизменной. В момент образования машины времени (преобразования пространственной червоточины во временную) между ее концами появляется замкнутый световой луч. Его энергия не достигает бесконечности, несмотря на бесконечное число проходов, из-за расфокусировки света [16]. Иная ситуация однако возникает в квазиклассической теории с «вакуумными флуктуациями» радиационного поля [14]. Проходя бесконечное число раз через червоточину и складываясь, эти флюктуации достигают бесконечной энергии, которая разрушит любого путешественника.

Однако ситуация в квантовой гравитации иная. Поскольку квантовые флуктуации содержат большие энергии, когда они возникают на коротких дистанциях, возможно найти настолько малое расстояние, на котором энергия флуктуации будет достаточно большой для формирования крошечной черной дыры, при этом горизонт черной дыры будет того же размера, как и это маленькое расстояние. Пространство-время не способно оставаться однородным на таких коротких дистанциях. Этот механизм обеспечивает естественную «блокировку» образованию сингулярных флюктуаций, ограничивая их по размеру – «максимальная энергия в минимальных размерах» [16].

Детальные расчеты квантовой гравитации показывают [15], что эта «блокировка» образованию сингулярных флюктуаций обеспечивает для макроскопических тел очень малую, но не нулевую вероятность беспрепятственного прохождения временной «червоточины». Как предотвратить в этой ситуации «парадокс дедушки»? Тут нам удобно воспользоваться языком многомировой интерпретации квантовой механики. Чтобы предотвратить парадокс путешественник должен попасть в параллельный мир, где он может беспрепятственно «убить дедушку» не нарушая принципа причинности. Такой параллельный мир будет квантово интерферировать с миром «неубитого дедушки», где наблюдателю не удалось пройти временную червоточину. Однако амплитуда вероятности такого мира будет крайне мала. Может ли наблюдатель в мире, где «дедушку не убили» обнаружить альтернативный мир хотя бы в принципе, используя квантовые корреляции между мирами? Подобно «парадоксу Шредингеровского кота» сделать он это не сможет. И причины те же, что и в обычной квантовой механике [2]. Проявления квантовых корреляций на тот момент, когда их величина существенна, невозможно зарегистрировать из-за «стирания памяти» наблюдателя. Таким образом, попадание в параллельный мир квантовой механики ничем экспериментально неотличимо от перестройки червоточины и попадания в «параллельный мир» общей теории относительности [3, 17]. Это значит, что с точки зрения внешнего реального макроскопического наблюдателя ситуация, когда путешественник погиб в червоточине или попал в «иной мир», экспериментально неотличимы. Это эквивалентно ситуации, когда путешественник падает в черную дыру: нам не известно будет ли он раздавлен в сингулярности или попадет в «иной мир» через белую дыру. [18] (Хотя для самого путешественника эта разница наблюдаема и существенна. Но он унесет свои все эти свои наблюдения с собой в «иной мир».) Мы видим, что как и в случае «информационного парадокса», разница между квантовой и квазиклассическими теориями для макроскопических объектов экспериментально не наблюдаема для макроскопического наблюдателя, не путешествующего во временной червоточине.

## 4. Черные звезды.

Недавно появилась интересная теория «черных звезд» [4]. Обычно коллапс черной дыры рассматривается как быстрый процесс. Однако нам не так хорошо известны состояния вещества при высоких давлениях. Мы знаем, что на пути к коллапсу возможно образование белых карликов, нейтронных звезд. Эти промежуточные стадии делают коллапс не лавинообразным, а постепенным. Возможно, на пути к коллапсу будут появляться кварковые звезды и иные промежуточные состояния, которые сделают этот процесс плавным. Для классической гравитации это несущественно. Но в квазиклассической гравитации показано, что при медленном сжатии квантовые флуктуации у поверхности помешают материалу звезды сколлапсировать в сингулярность и стать черной дырой. Снаружи такой объект будет подобен черной дыре, но внутри будет от нее отличен, не содержа сингулярность и сохраняя всю информацию. Он позволит путешественнику входить за поверхность звезды и выходить из нее. Следует отметить, что против такой картины имеется существенное возражение.

Насколько стабильна такая конструкция звезды внешнему возмущению, вносимому путешественником? И насколько устойчив сам путешественник? Устойчива ли она? Путешественник – это макроскопическое тело. Проникнув в черную звезду, он скачком увеличит ее массу. Это приведет либо к её схлопыванию в черную дыру. Если же процесс снова пойдет «постепенно» то путешественник «растворится» в звезде и не сможет из нее выйти. Таким образом, видится, что разница между черной звездой и черной дырой не наблюдаема экспериментально. А, значит, разница между этими объектами существует лишь на бумаге – в идеальной динамике.

## 5. Новая теория гравитации Пенроуза.

В своих замечательных книжках [6-7] Пенроуз дает свое замечательное предвидение новой грядущей теории гравитации. В этой теории:

1) В отличие от обычной квантовой механики редукция волнового пакета – фундаментальное свойство теории.
2) Это редукция неразрывно связана с явлением гравитации.
3) Редукция приводит не только к вероятностным закономерностям – но она может приводить и к более сложному «неопределенному» поведению систем.
4) В отличие от замечательных когерентных квантовых систем, классические хаотические неустойчивые системы подвергаются уничижающей критике, как чисто абстрактные модели, бесполезные для понимания реальности. Они не имеют никакого отношения к описанным выше «неопределенным» системам, которые могут быть лишь чисто квантовыми.

Следует заметить, что для получения всех этих вещей нам нет нужды в новых теориях. Учет неизбежного гравитационного взаимодействия макроскопического реального наблюдателя (плюс учет его термодинамической стрелы времени) с необходимостью ведет ко всем этим результатам в рамках любой уже существующей теории квантовой гравитации. Кроме того, классические хаотические неустойчивые системы обладают всеми свойствами квантовых. Для любого «чисто квантового эффекта» всегда можно найти такой классический аналог (Приложение A [2]). А именно:

1) Мы видели из описанного выше, что неизбежное гравитационного взаимодействия макроскопического реального наблюдателя с наблюдаемой системой неизбежно делает ее эволюцию неунитарной. Разница между унитарной и неунитарной теорией существует лишь на бумаге и экспериментально не наблюдаема в квантовой теории гравитации.
2) По изложенным выше причинам именно гравитация ведет в уже существующей квантовой теории гравитации к неизбежной редукции и неунитарности. Более того, для макроскопических объектов желаемая неунитарная теория, обладающая всеми желаемыми свойствами, уже существует – это квазиклассическая теория гравитации.
3) Поведение многих макроскопических тел, несмотря на неунитарность, можно достаточно полно описать набором макропараметров и законами их эволюции. Имеются, однако, *непредсказуемые* системы, чьё поведение во всей их полноте затруднительно описать даже вероятностными законами.
   Например, квантовые компьютеры. Для человека, запустившего такой квантовый компьютер и знающего его начальное состояние, его поведение полностью предсказуемо. Однако для человека, не присутствующего при запуске, его поведение «неопределенно». Более того, попытка пронаблюдать внутренне состояние квантового компьютера приведет к нарушению его нормальной работы.
   В случае квантовой гравитации даже запустивший квантовый компьютер человек не сможет 100% предсказать результат – неизбежное гравитационное взаимодействие между ним и квантовым компьютером сделает такое предсказание невозможным. Таким образом, «непредсказуемость, отличная от вероятностной» становится фундаментальным свойством любой квантовой теории гравитации при учете неизбежного взаимодействия с наблюдателем.
4) Неустойчивые классические системы во многом напоминают по своим свойствам квантовые (Приложение A [2]). Более того, созданы математические модели классических аналогов квантовых компьютеров [19]. Для моделирования парадоксальных свойств живых объектов, напоминающих квантовые компьютеры, нам не обязательно нужна квантовая механика. Можно обойтись и классическими неустойчивыми системами [20].

   Подводя итог, мы видим, что все пожелания Пенроуза воплотимы уже в рамках существующей парадигмы и не требуют никаких новых фундаментальных теорий. Более того, все свойства объектов описываются макропараметрами (чтобы исключить влияние макроскопического наблюдателя), что неизбежно ведет к ненаблюдаемости слишком малых интервалов времени и пространства. А значит можно построить их наблюдаемую динамику на основе «дискретной модели пространства-времени». Но такая модель уже не будет новой теорией – для любого макроскопического наблюдателя она будет экспериментально неотличима от уже существующей квантовой теории гравитации.

## 6. Антропный принцип в квантовой теории гравитации.

Количество возможных вакуумных состояний, возникающих в квантовой теории гравитации, достигает огромного количества. Для отбора подходящих вакуумов обычно предлагается антропный принцип [5] - эволюция системы должна приводит к появлению наблюдателя, способного ее понимать и изучать. Подобная постановка вопроса носит слишком философский характер, затрудняющий ее практическое использование. В данной работе мы можем сформулировать вполне четкие физические принципы, по сути эквивалентные антропному принципу:

Начальное состояние Вселенной должно приводить к формированию материи в виде набора многих макроскопических термодинамически неравновесных тел, слабо взаимодействующих между собой. Эти тела должны иметь энтропию и температуру. Они должны иметь собственную термодинамическую стрелу времени. Малое локальное взаимодействие между телами должно приводить к синхронизации их стрел времени. Хотя эти тела состоят из многих частиц и обладают огромным набором микропарпаметров, эволюция этих тел должна хорошо описываться набором макропараметров, за исключением редких положений неустойчивости.

Однако эти неустойчивые состояния играют большую роль, служа основой для возникновения самого наблюдателя Вселенной. Должны существовать неустойчивые глобальные корреляции между частями мира, так и неравновесные системы с поддерживаемыми локальными внутренними корреляциями, из которых и образуются сами наблюдатели.

Глобальной «целью» диссипативных систем с локальными корреляциями (в том числе и живых систем) является (а) минимизация их собственной энтропии (б) стимуляция глобальной полной системы к скорейший возврату Пуанкаре в исходное низкоэнтропийное состояние.

Из вышесказанного следует один важный вывод. Чтобы получить вышеописанную ситуацию, начальное состояние Вселенной должно неизбежно быть высокоупорядоченным и низкоэнтропийным.

Т.е., коротко говоря, эволюция должна приводить к миру, допускающему описание в форме термодинамики [1-3, 21-23]. Только в такой среде может появиться наблюдатель, способный изучать этот мир.

## 6. Выводы

Мы видим, что информационный парадокс и парадокс дедушки разрешаются в квантовой гравитационной теории аналогично общей теории относительности. Это делается путем рассмотрения взаимодействия систем с реальным неравновесным макроскопическим наблюдателем. Более того, этот подход аналогично обычной квантовой теории, позволяет разрешить проблему редукции. Это редукция в квантовой гравитации становится фундаментальным свойством теории, в отличие от обычной квантовой механики. Такой подход позволяет рассмотреть и другие сложные вопросы квантовой гравитации – антропный принцип, черные звезды.

## Благодарности

## Библиография